\documentclass[superscriptaddress]{article}
\usepackage{authblk}
\usepackage[hidelinks,colorlinks=true,allcolors=blue]{hyperref}
\usepackage[hmargin=0.5in, vmargin=0.5in]{geometry}
\usepackage{indentfirst}
\usepackage{amssymb}
\usepackage{amsmath}
\usepackage{graphicx}
\usepackage{physics}
\usepackage{braket}
\usepackage{caption}
\usepackage{subcaption}
\usepackage{multirow}
\usepackage{bm}
\usepackage[square,numbers]{natbib}

\usepackage{tikz}
\usepackage{xr}
\usepackage{siunitx}

\title{\Large \textbf{Measurements of electronic band structure in \texorpdfstring{CeCoGe\textsubscript{3}}{CeCoGe3} by angle-resolved photoemission spectroscopy}}

\author{Robert Prater $^{1,2}$, Mingkun Chen $^{1}$, Matthew Staab $^{1}$, Sudheer Sreedhar $^{1}$, Journey Byland $^{1}$, Zihao Shen $^{1}$, Sergey Y. Savrasov $^{1}$, Valentin Taufour$^{1}$, Vsevolod Ivanov$^{3,4,5}$,  and Inna Vishik $^{1,2}$*}

\affil[1]{\small Department of Physics and Astronomy, University of California, Davis, CA 95616, USA}
\affil[2]{Materials Sciences Division, Lawrence Berkeley National Lab, Berkeley, California 94720, USA}
\affil[3]{Department of Physics, Virginia Tech, Blacksburg, Virginia 24061, USA; vivanov@vt.edu}
\affil[4]{Virginia Tech National Security Institute, Blacksburg, Virginia 24060, USA}
\affil[5]{Virginia Tech Center for Quantum Information Science and Engineering, Blacksburg, Virginia 24061, USA}
\begin{document}
\maketitle
\section*{Abstract} 

We report a comprehensive study of the electronic structure of CeCoGe$_3$ throughout the entire Brillouin zone in the non-magnetic regime using angle-resolved photoemission spectroscopy (ARPES). The electronic structure agrees in large part with first principles calculations, including predicted topological nodal lines. Two new features in the band structure are also observed: a surface state and folded bands, the latter which is argued to originate from a unit cell reconstruction.



\section{Introduction}

The field of unconventional superconductivity was invigorated by the discovery of superconducting phases in materials lacking inversion symmetry. The lack of parity as a good quantum number permits the existence of anti-symmetric spin-orbit coupling interactions that can lead to unexpected behavior of the superconducting state. These interactions result in superconducting pairing that is a mixture of singlet and triplet states, as well as a complex structure of the superconducting gap that can include point nodes and line nodes \cite{Sigrist2009, Smidman2017, Fischer2023}.

Electronic correlations and band topology can further add richness to the potential pairing mechanisms and phenomenologies in superconductors that lack inversion symmetry, and it was recently demonstrated \cite{ivanov2021renormalized} that these components coexist in non-centrosymmetric heavy fermion materials with chemical formula CeTX$_3$ (T$=$transition metal, X$=$ Si or Ge), with the tetragonal BaNiSn$_3$ crystal structure.  Collectively, these compounds span the full Doniach phase diagram from localized magnetism to heavy fermion physics without magnetic order \cite{Weng:Ce_HF_2016,Rai_2022}.  CeCoGe$_3$, the compound of interest in this manuscript, is in the intermediate regime, featuring enhanced carrier mass and magnetic ordering \cite{Eom:quantumCritical_CeCoGe,Pecharsky:MagneticProperties1993,Thamizhavel:QO_CeCoGe3}. This competition between magnetism and Kondo physics was demonstrated via temperature-dependence of spectral weight measured by ARPES \cite{Li:PhotoemissionSignature2023}. Recent studies have also investigated predicted Weyl nodes near the Fermi energy ($E_F$)  \cite{ivanov2021renormalized, Furuashi:DopingInducedAHE, moya2025measuringhalleffecthysteretic}.

CeCoGe$_3$ is susceptible to superconductivity under hydrostatic pressure  \cite{Settai:Pressure_SC_CeCoGe3} and can be driven to the magnetic quantum phase transition  via chemical substitution \cite{Skokowski_2019, Eom:quantumCritical_CeCoGe}. In the superconducting state, CeCoGe$_3$ and related compounds are distinguished by extremely high upper critical magnetic fields along the $c$-axis \cite{Kimura:Highc2CeRhSi3_2007,M_asson_2010}. This, together with the broken inversion symmetry, topological band structure, and the appearance of superconductivity at the endpoint of magnetic order has been cited in proposal of unconventional, possibly triplet, superconductivity in this and related compounds \cite{Kawai_2008}.  Experimental fermiology is one of the ingredients for evaluating unconventional superconducting mechanisms.

Here we report a comprehensive ARPES study of the non-magnetic electronic structure of CeCoGe$_3$ throughout the three dimensional (3D) Brillouin zone. There is overall agreement with first-principles calculations that assume localized $f$ electrons.  However, two additional features are seen experimentally: a two dimensional (2D) surface-like band and band folding that appears to be most consistent with a unit-cell doubling. 


\section{Materials and Methods}
Single crystals of CeCoGe$_3$ were synthesized by the solution growth method \cite{Thamizhavel:UniqueMagneticCeCoGe3}. First, the stoichiometric composition of CeCoGe$_{3}$ was arc-melted, flipped upside-down and arc-melted again multiple times to ensure a homogeneous mixture. It was then combined with bismuth in a ratio of Ce$_{8}$Co$_{8}$Ge$_{24}$Bi$_{60}$. The entire mixture was initially heated to $1150$ \textdegree C within $6$ hours, followed by a dwell time of $72$ hours. The temperature was then slowly decreased to $750$ \textdegree C over $90$ hours, after which the remaining molten bismuth flux was removed by centrifugation.

ARPES and XPS measurements were performed at the MERLIN ARPES endstation (Beamline 4.0.3) at the Advanced Lightsource and beamline 5.2 at the Stanford Synchrotron Radiation Lightsource (SSRL) using photon energies between 22 and 150eV. Beam spot size at SSRL was $\approx 15 \times 8 $ \SI{}{\micro\meter}   and the analyzer energy resolution was $\approx 18$ meV.   MERLIN data were taken with spot size $\approx 50 \times 70$ \SI{}{\micro\meter} and energy resolution $\approx 25$ meV. All photoemission spectra were collected at 30K, above any of the magnetic ordering transitions \cite{Thamizhavel:UniqueMagneticCeCoGe3}.

First-principles calculations were carried out using the full potential linear muffin-tin orbital method including spin-orbit coupling \cite{lmtart}. The effect of Ce$4f$-orbital electronic correlations are captured within the Local Density Approximation + Gutzwiller (LDA+G) formalism \cite{ldaG, sav_ldaG}. The local crystal field effects on the Ce $4f^1_{7/2}$ and Ce $4f^1_{5/2}$ multiplets are captured using a band-dependent double counting scheme \cite{ivanov2021renormalized}. Topological features within the Brillouin zone (BZ) were identified recursively using a Berry curvature link-variable approach \cite{mmm}.

\section{Results}

\begin{figure}[h!]
\centering
\includegraphics[width=10.0 cm]{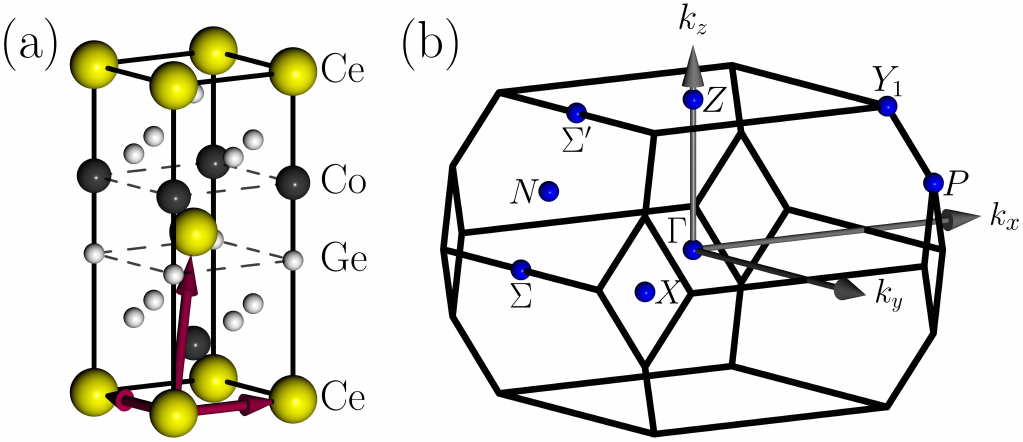}
\caption[Crystal structure and Brillouin zone.]{CeCoGe$_3$ (a) crystal structure (conventional unit cell).  Arrows indicate primitive lattice vectors. (b) Brillouin Zone with high symmetry points labeled. }
\label{CrystalStructureAndBZ}
\end{figure}   
\unskip

CeCoGe$_3$ crystallizes in a non-centrosymmetric tetragonal crystal structure with the space group I4mm (No. 107); the structural (conventional) unit cell is shown in Fig. \ref{CrystalStructureAndBZ}(a), but the primitive cell is used to define the BZ and for performing calculations. The conventional unit cell can be viewed as a layered structure with Ce, Co, and two adjacent non-equivalent Ge layers.  The 3D BZ is shown in Fig. \ref{CrystalStructureAndBZ}(b), with high symmetry points marked.  This work explores electronic structure throughout the 3D BZ.  Throughout the text, energy vs. momentum high-symmetry cuts are taken either parallel to $\Gamma-\Sigma$ or parallel to $\Gamma-X$ at different values of $k_z$. The $\Gamma$ plane, $Z$ plane, and $N-P$ plane refer to planes through these high symmetry points and parallel to $k_x-k_y$.  $\overline{\Gamma Z}$ refers to the line between $\Gamma$ and $Z$ along the $k_z$ axis. 

\begin{figure}[h!]
\centering
\includegraphics[width=10.0 cm]{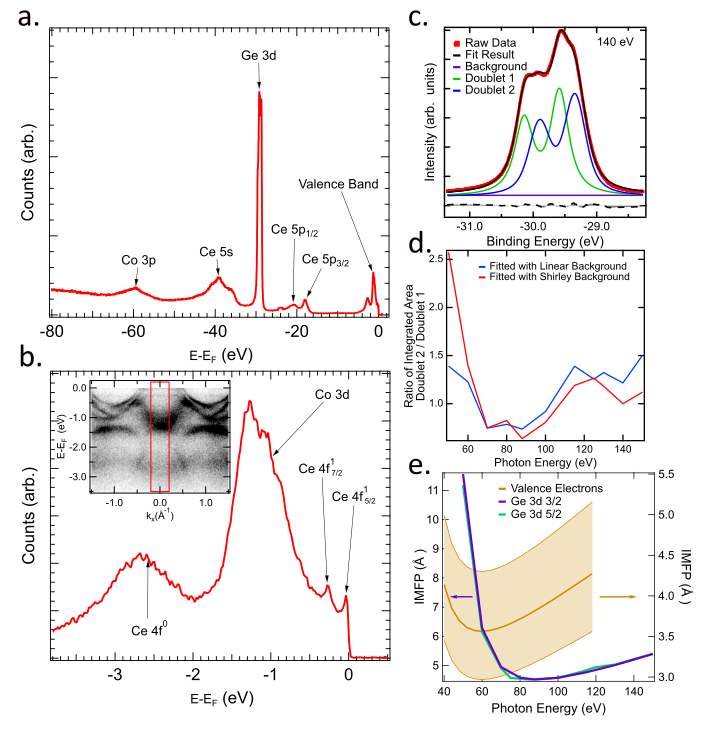}
\caption{Core level spectroscopy. (a) Core level survey. (b) Momentum-integrated valence band and localized contributions from the Ce $4f$ derived states.  Inset shows original energy vs momentum cut, together with integration window.  Panels (a)-(b) taken at 125 eV. (c) Ge $3d$ level taken at 140 eV, together with fitting to two doublets. Dashed line is residual of fit.  (d) photon energy dependence of intensity ratio of two fitted doublets (e) calculated IMFP as a function of photon energy for Ge $3d_{1/2}$ and $3d_{3/2}$ (left axis) as well as valence electrons (right axis). 
\label{XPS_fig}}
\end{figure} 

Fig. \ref{XPS_fig}(a) shows a survey XPS spectrum.  All elements of the compound are present and no extraneous peaks are observed, which indicates a cleave without contamination from Bi flux.  An example spectrum for a sample with residual flux is included in the supplementary materials; samples such as these tend to yield poor quality ARPES spectra. Fig. \ref{XPS_fig}(b) highlights the valence bands, with an emphasis on Ce levels. In Fig. \ref{XPS_fig}(c), we focus on the Ge-$3d$ states, which clearly show at least four peaks.  These spectra were measured in the 50-150 eV photon energy range and fitted using a background plus two Voigt-profile spin–orbit-split doublets, from which the integrated area of each component was extracted. Both linear and Shirley background functions were tested in the fitting (See supplementary materials). The results of this fit are summarized in Fig. \ref{XPS_fig}(d), which plots the area ratio of the two doublets. Though the results are background-dependent at low photon energy, the doublet at higher binding energy (doublet 1) is strongest at a photon energy of $\approx 90$ eV, where the inelastic mean free path (IMFP) for Ge $3d$ electrons is the smallest (Fig. \ref{XPS_fig}(e)).  For the Ge$-3d$ core levels, IMFPs were calculated with SESSA v2.2.2 \cite{sessa}, yielding the two core-level curves (3d$_{5/2}$ and 3d$_{3/2}$), which differ slightly due to their different kinetic energies at a given photon energy. For valence band photoelectrons (Fig. \ref{XPS_fig}(e)), a reference IMFP was computed using the semi-empirical Gries G1 expression \cite{Gries}. In this model, the effective electron number Z$^{*}$ represents the number of electrons that most strongly contribute to inelastic scattering. The envelope spans Gries-G1 results using $Z^{*}=Z_{\mathrm{eff}} (Z_{\mathrm{eff}}=(\sum_i f_i Z_i^{2.94})^{1/2.94}=43.44)$ \cite{effective_Z} and Z$^{*}=30$ (valence-electron count\cite{TPP2m}), while the central curve uses the direct atomic-number average. IMFP values for Ge-$3d$ levels and valence bands are calculated only over the measurement range of these quantities in this manuscript.

\begin{figure}[h!]
\centering
\includegraphics[width=18 cm]{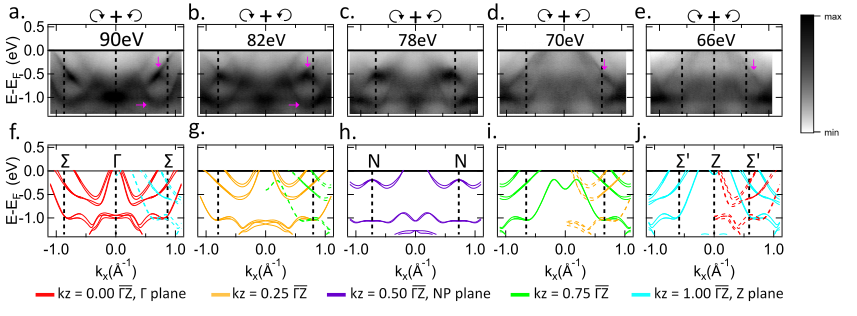}
\caption{ Cuts parallel to $\Gamma-\Sigma$ at different $k_z$. (a)-(e) Energy vs momentum cuts at photon energies indicated in each panel. Spectra collected with  RCP+LCP light. Pink vertical arrows point to `x-shaped' feature and horizontal pink arrow points to dispersing feature not directly captured in calculation. (f)-(j) LDA+G calculations from Ref. \cite{ivanov2021renormalized} corresponding to each data panel above.  Calculations have been shifted to lower binding energy (towards $E_F$) by 180 meV for better agreement with data. Vertical dashed lines denote high symmetry points in (f),(h) and (j), and the edge of the BZ at that value of $k_z$ otherwise. Right half of panels (f),(g),(i),(j) shows overlay of calculation from half a BZ away along $k_z$, with color of these dashed lines corresponding to legend labels below.  Color bar on right applies to all ARPES image plots in manuscript.
 \label{fig3}}
\end{figure}   
\unskip

ARPES spectra were collected throughout the BZ, and most of the detailed spectra focus near planes cutting through the $\Gamma-Z$ line ($\overline{\Gamma Z}$) at 5 roughly equidistant positions:  $\Gamma$ plane ($k_z = 0.0$ $\overline{\Gamma Z}$), $k_z = 0.25$ $\overline{\Gamma Z}$, $0.5$ $\overline{\Gamma Z}$ (also called $N-P$ plane), $0.75$ $\overline{\Gamma Z}$, and the $Z$ plane ($1.0$ $\overline{\Gamma Z}$).  Fig. \ref{fig3} shows high symmetry cuts parallel to $\Gamma-\Sigma$ at these values of $k_z$. The spectra shown are a sum of spectra collected with left circularly polarized (LCP) and right circularly polarized (RCP) light at the photon energy indicated in each panel. These energy vs momentum spectra are energy distribution curve (EDC) normalized by taking each momentum channel and dividing it by the integrated area of that channel. The spectra tend to have weaker intensity close to $E_F$, which is also seen in previously published data \cite{Li:PhotoemissionSignature2023}.  Qualitatively, our spectra show good agreement with LDA+G calculations in Ref. \cite{ivanov2021renormalized}, albeit with several differences.  First, experimental spectra show a rigid shift of $\approx 180$ meV relative to the calculation.  All calculations shown in this manuscript employ this rigid shift relative to Ref. \cite{ivanov2021renormalized}.  We note that a rigid shift does not give perfect agreement, especially in the NP plane (Fig. \ref{fig3}(c),(h)).  Additionally, there are measured bands that are absent in the calculation.  When these additional bands cross the predicted bands, they form `x-shaped' features at low binding energy, and some examples are marked by vertical arrows in Fig. \ref{fig3}(a)-(e).  Additional extra features are marked by horizontal arrows in some panels. At most planes in the BZ, the position and dispersion of these addition features is captured by calculated bands half the $\overline{\Gamma Z}$ distance; for example extra bands near the $\Gamma$ plane agree well with bands near the $Z$ plane. Examples of these offset bands are plotted in dashed lines in Fig. \ref{fig3}(f,g,i,j). Panel (h) corresponds to the NP plane, and half a BZ away in $k_z$ is still the NP plane.  

\begin{figure}[h!]
\centering
\includegraphics[width=16.0 cm]{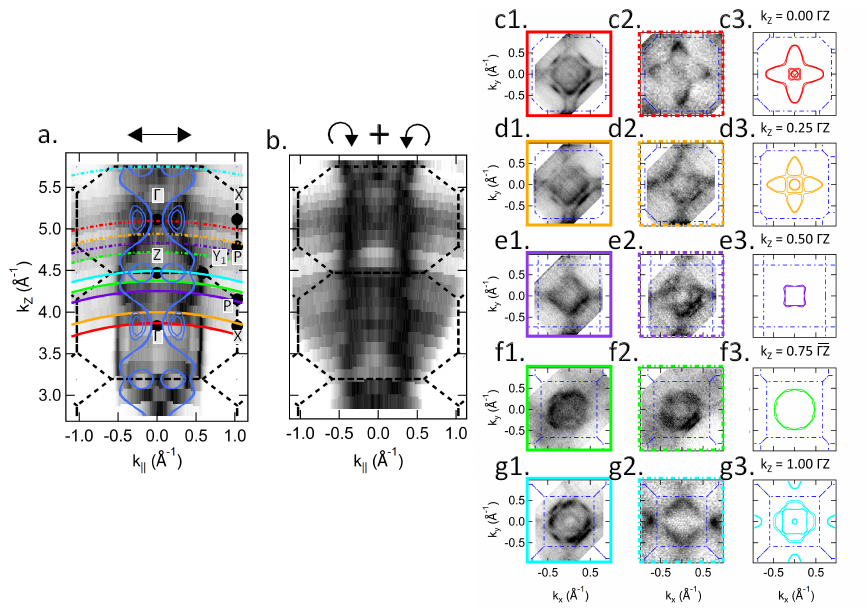}
\caption{Constant energy maps at $E_F$. (a) Fermi surface map in $\Gamma-X-Z$ plane taken with LH light in photon energy range 26-118 eV.  Yellow dots mark high symmetry points.  Solid and dashed lines correspond to locations of maps in panels (c)-(g). (b) Fermi surface map in same momentum space region as (a) but taken with LCP+RCP light. (c)-(g) Constant energy maps in $k_x$-$k_y$ plane, taken with LH light. Data taken at select photon energies to closely match to the following planes along $\overline{\Gamma Z}$:  $\Gamma$ plane, $k_z = 0.25$ $\overline{\Gamma Z}, 0.5$ $\overline{\Gamma Z}, 0.75$ $\overline{\Gamma Z},$ and the $Z$ plane. Column 1 is taken at lower photon energies (solid lines in (a)).  Second column is taken at higher photon energy (dashed lines in (a)), but closely matching locations in BZ. Third column is calculated Fermi surfaces for the  $k_z$ values probed by data.
\label{fig4}}
\end{figure}

Fig. \ref{fig4} explores constant energy maps at $E_F$ (Fermi surface maps) both along $k_z$ and perpendicular to $k_z$. All maps are generated with a $60$ meV integration window centered around $E_F$. Fig. \ref{fig4}(a)-(b) were collected in the $\Gamma-X-Z$ plane, where photon energy was tuned between 26 and 118 eV to access different values of $k_z$.  An inner potential value of 15.5 eV was used, as determined from periodicity of the spectra and from matching Fermi surface maps to calculations \cite{Damascelli_2004_inner_potential}.  Different polarizations were used in panels (a) (Linear horizontal, LH) and (b)(RCP+LCP) to capture different band features. Fig. \ref{fig4} (a) and (b) were taken on the same cleave. Panel (a) shows some features that agree with overlaid calculated Fermi surfaces, such as near the BZ boundary at $Z$, but other experimental features not captured in calculation, such as near $\Gamma$ where the measured Fermi surface is further away from the zone center than the calculation. Additionally, there are `vertical' features prominent in the bottom half of Fig. \ref{fig4} (a).  Because these features are very sharp and do not disperse as a function of $k_z$, they are presumed to be surface states.  These features are not visible in panel (b), which was taken with a different polarization.  In those data, the Fermi surface in the $\Gamma-X-Z$ plane appears to have a periodicity twice that of the BZ in $k_z$.  

In Fig. \ref{fig4}(c)-(g) we examine Fermi surface maps as a function of $k_x$-$k_y$ for different values of $k_z$ spaced roughly evenly along the $\overline{\Gamma Z}$. All of these maps are collected with LH polarization.  For each sampled plane of the BZ, Fermi surface maps collected at different photon energies are shown in columns 1 and 2 in order to disambiguate effects of matrix elements and differing surface sensitivity. Calculated Fermi surfaces are shown in column 3.  While the calculations are exactly at the planes indicated in the figure, data are collected along the trajectories indicated in (a) which are not planar and not located exactly equidistantly along the $k_z$ axis. The surface bands discussed for Fig. \ref{fig4}(a) are seen most clearly as sharp quarter-circles furthest from the center of the zone in (c1), (d1), (e1), (e2), and (g1).  Panels (c2), (d2), and (g2) agree reasonably well with their accompanying calculation.  Meanwhile some of the other measurements agree better with calculations elsewhere in the BZ: (c1), ignoring the surface band, has a diamond-shape surrounding a square similar to (g3); (d1) has a prominent diamond-shaped feature lacking in (d3); (e2) has bands additional to the predicted square; (f2) has bands other than the predicted circle.  

\unskip

\begin{figure}[h!]
\centering
\includegraphics[width=14.0 cm]{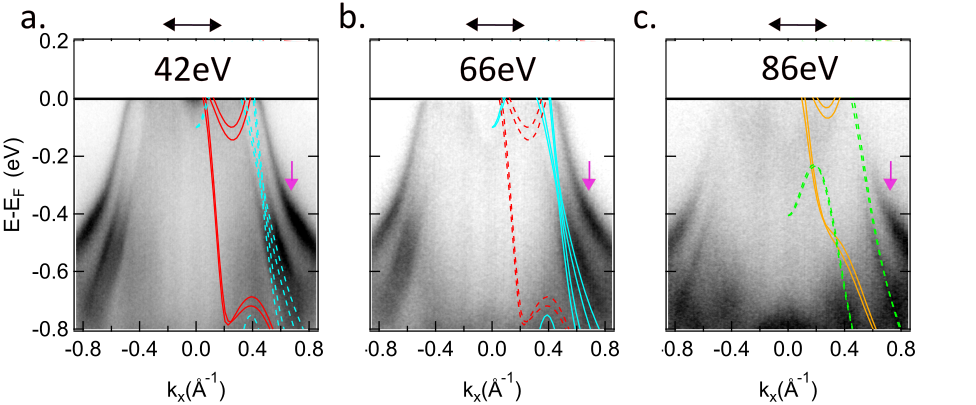}
\caption{Surface-like features in CeCoGe$_3$.  All cuts are taken parallel to $\Gamma-\Sigma$ at different photon energies with LH polarization.  Calculations are shown in measured plane (solid) and half a BZ away in $k_z$ (dashed). Pink arrow marks one instance of this surface band in each panel. (a) 42 eV (b) 66 eV (near Z plane) (c) 86 eV. 
}
\label{fig5}
\end{figure}

In Fig. \ref{fig5} we provide more details about the surface-like bands prominent for measurements with LH polarization. The dispersing bands closer to the zone center agree well with calculated bands.  However, the additional band marked by the pink arrow differs substantially, especially at deeper binding energy.  These extra features have a Fermi crossing around $k_{||}=0.44$ \AA{}, a band bottom $\approx -0.6$ eV, and the Fermi velocity is  $\approx2$ eV$\cdot$\AA. At some photon energies, these surface bands have weak intensity at $E_F$ and stronger intensity at deeper binding energy (e.g. Fig.\ref{fig5}(c)), while others show strong intensity at $E_F$ (e.g. Fig. \ref{fig5}(a),(b)), contributing to the sharp 'vertical' features in Fig. \ref{fig4} (a).

\begin{figure}[h!]
\centering
\includegraphics[width=14.0 cm]{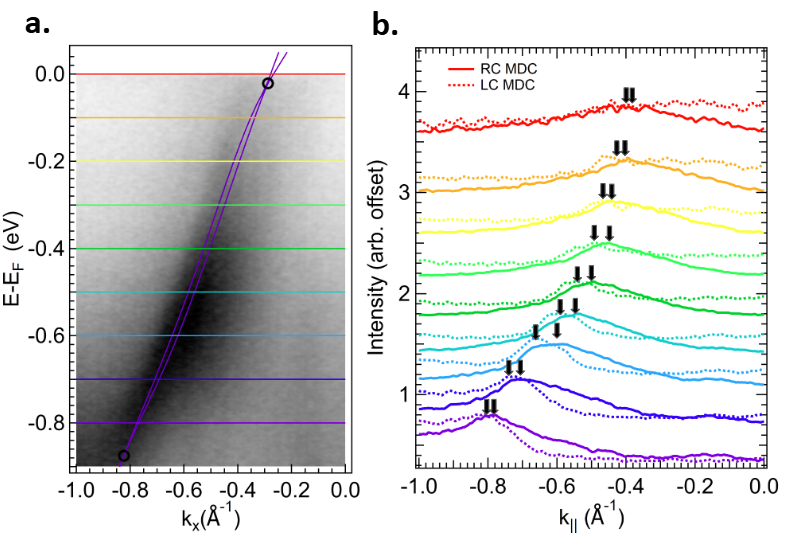}
\caption{Evidence for nodal lines. (a), spectra taken from the N-P plane ($k_z = 0.5\overline{\Gamma Z}$) with LCP light at 78eV. Purple: overlay of calculated bands with black circles indicating the crossing point which forms the nodal lines. (b), MDCs at energies indicated in (a), collected with RCP (solid) and LCP (dashed) light. Arrows are guide-to-eye for MDC peak positions.\label{fig6}}
\end{figure}

In the non-magnetic state where the present measurements are performed, calculations have predicted that CeCoGe$_3$ is a topological metal, featuring multiple Weyl points and Weyl nodal lines throughout the BZ \cite{ivanov2021renormalized}. Fig. \ref{fig6} explores some evidence for these features, which are challenging to observe because the band splitting in the vicinity of the nodes tends to be small. Different momentum distribution curves (MDCs) are examined with LCP and RCP light, and the peak position is identified as the band position.  Fig. \ref{fig6}(b) shows the splitting and re-joining of bands between two predicted topological crossings, where each band is highlighted with a different polarization.

\unskip

\unskip

\section{Discussion}



The agreement between LDA+G in Ref. \cite{ivanov2021renormalized} and ARPES data is quite good throughout most of the BZ, as long as only weak hybridization with $4f$-electrons is considered. This localized Ce $4f$ electron character is supported by the Ce $4f^0$ peak appearing at binding energies deeper than 2 eV as well as a relatively high intensity of $4f^1_{7/2}$ as compared to $4f^1_{5/2}$ \cite{Chen:LocalizedToItinerant_2018}. The strongest  indicator in the present data is the overall energy shift between calculations involving hybridized $4f$ electrons (Ref. \cite{ivanov2021renormalized}) and measured band structure (Fig. \ref{fig3}).  This shift is well captured by performing calculations absent hybridized Ce electrons (supplementary materials and Ref. \cite{Li:PhotoemissionSignature2023}).  Weak hybridization between $f$ electrons and conduction elections was also reported in studies on polycrystalline specimens \cite{Skokowski_2019_XPS}.     

This compound is predicted to have topological band features in the non-magnetic state that are interesting in the context of this material's moderate electronic correlations and tendency towards superconductivity under pressure.  There is evidence for these nodal features via slight differences in dispersion for spectra taken with different polarization, in a manner that matches predicted energy-dependent splitting and re-joining of bands (Fig. \ref{fig6}).  The example shown in that figure has a Weyl crossing near $E_F$, with relevance to low energy phenomena like transport and superconducting pairing.  When evaluating which Weyl crossings are relevant to transport, the rigid shift between measured band structure and predictions in Ref. \cite{ivanov2021renormalized} needs to be considered.  

The crystal structure of CeCoGe$_3$ has two in-equivalent Ge positions \cite{Rogalev_2021}, and it is thought that this compound cleaves with a Ge termination \cite{Li:PhotoemissionSignature2023}.  Both of these statements could be consistent with the two sets of Ge-$3d$ doublets that are shown in Fig. \ref{XPS_fig}(c), and the relative intensity of the two doublets as a function of photon energy helps disambiguate between these scenarios. When measured at around 90 eV photon energy (Fig. \ref{XPS_fig}(d)), the relative intensity of the higher-binding-energy doublet (doublet 1) is highest.  Although the detailed energy dependence is influenced by the choice of background at low photon energy, the general photon-energy dependence is robust. This energy is also near the value where the IMFP is minimum for Ge-$3d$ electrons.  Together this suggests, assuming the observation is primarily from initial-state effects, that the doublet at higher binding energy originates from the surface, and the relatively large ($\approx 0.5$ eV) chemical shift of the surface Ge species can be concomitant with the appearance of surface states or structural reconstruction.  

One new experimental feature in the present data is a surface state that appears around the corners of the BZ. This feature manifests as dispersionless states as a function of photon energy in Fig. \ref{fig4}(a)) and Fig. \ref{fig5}, and these bands are extremely polarization dependent, being almost entirely absent with circularly polarized light, but very strong with LH light.  We note that some LH spectra simultaneously show these surface states and reconstruction features from unit cell doubling (Fig. \ref{fig5}(b),(c)), indicating that they are not mutually exclusive.  These surface states may be associated with topological band features, or they may be topologically trivial surface bands related to the strong surface effects exhibited by terminating Ge atoms (Fig. \ref{XPS_fig}).

The other new experimental observation is the additional band features shown in Fig. \ref{fig3} and \ref{fig4}, and these extra features show strong similarity with bands that are half a BZ away in $k_z$ for most cuts where they are observed.  This suggests a band folding feature from some incipient order.  The choice of folded bands (dashed) assumes an ordering vector of $\mathbf{q}=(0,0,1/2)$, but the observed spectra could also be consistent with a unit cell doubling in $a$ or $b$ because of how the BZs are tiled.  However, the former reconstruction is tentatively assumed because known orders in this material are along $c$, though present measurements at 30K were well above any of the magnetic ordering temperatures \cite{Pecharsky:MagneticProperties1993,Thamizhavel:UniqueMagneticCeCoGe3,Smidman:NeutronMuon2013}.  The presumed unit cell doubling is consistent with ground state bulk magnetic order.  Below $T_{N1}=21K$, the magnetic ordering vector is $\mathbf{q}=(0,0,2/3)$, and this is closest to our measurement temperature. We do not fully rule out this ordering vector, but our data slightly favor $\mathbf{q}=(0,0,1/2)$ because of strong mixing of $\Gamma$-plane and $Z$-plane spectra, the $\overline{\Gamma Z}/2$ periodicity of FS maps in Fig.\ref{fig4}(b), as well as large discrepancies between calculation and experiment in the $N-P$ plane, where bands would cross and hybridize under a doubled unit cell. The aspect of the data that more favors a $\mathbf{q}=(0,0,2/3)$ order is the observation that many of the FS maps outside of the $Z$ plane in Fig. \ref{fig4}(c)-(g) show strong $Z$-plane features of a square inside of a diamond (c1, d1, e2). A $\mathbf{q}=(0,0,2/3)$ order would repeat $Z$-plane spectra more times along $k_z$ even before considering the effects of perpendicular momentum resolution. Perpendicular momentum uncertainty/broadening originates from a short IMFP, and this can both broaden spectra and oversample extremal momenta of a dispersing band \cite{Bansil:k_perp_broadening_cuprate_2005,STROCOV2003}.  In the present experiments, momentum uncertainty is estimated to be $\Delta k_\perp=1/\text{IMFP}\approx 0.2$ \AA$^{-1}$, or $\approx 17\%$ of the BZ height. 



We propose that the folded bands originate from an additional reconstruction in the non-magnetic state due to bond distortion, charge disproportionation, or surface-stabilized magnetism, though our results do not directly distinguish between these proposals. This reconstruction may be primarily near the surface, as found in polar materials existing near magnetic phases \cite{Chikina_2014,Mazzola_2018}.  The present measurements focus on photon energies where ARPES is extremely surface sensitive, with an IMFP varying by $\approx 10\%$ around 4 \AA; information depth is typically considered to be 2-3$\times$ the IMFP.  A surface reconstruction would dominate our surface-sensitive measurements, but be imperceptible to bulk probes which have dominated existing literature.  However, our data do not rule out a weak bulk incipient order, and if such features are found, particularly under pressure, it may be relevant for the superconducting mechanism, both as a candidate wavevector for interactions related to pairing and as an origin of a significantly different fermiology.

\section{Conclusions}

We have mapped out the electronic structure of CeCoGe$_3$ over the entire 3D BZ in the non-magnetic state at 30K. The measured electronic structure agrees in large part with band structure calculations and quantum oscillation experiments, including predicted topological nodal lines. Two new features in the band structure \textemdash a surface state and folded bands tentatively attributed to a unit cell doubling stabilized by the surface \textemdash are also discussed.  



\vspace{6pt} 

\textbf{Funding:} This research was supported by Grant No. 2020067 from the United States-Israel Binational Science Foundation (BSF). M.C. acknowledges support by the National Science Foundation (US) Division of Materials Research Award. No. 2428464. Use of the Stanford Synchrotron Radiation Lightsource, SLAC National Accelerator Laboratory is supported by the U.S. Department of Energy, Office of Science, Office of Basic Energy Sciences under Contract No. DE-AC02-76SF00515. This research used resources of the Advanced Light Source, a U.S. Department of Energy Office of Science User Facility under Contract No. DE-AC02-05CH11231. VI acknowledges support from Virginia Tech startup funds. S.S acknowledges support from US DOE grant No. DE-SC0026106. 

\textbf{Data Availability Statement:} The data that support the findings of this manuscript will be openly available following publication.

\textbf{Acknowledgments:} The authors acknowledge helpful discussions with Jonathan Denlinger, Jamie Moya, and Ittai Sidilkover.

\end{document}


\setlength{\intextsep}{5pt}
\title{\Large
Supplementary Material: Measurements of electronic band structure in \texorpdfstring{CeCoGe\textsubscript{3}}{CeCoGe3} by Angle-resolved photoemission spectroscopy}


\author{Robert Prater $^{1,2}$, Mingkun Chen $^{1}$, Matthew Staab $^{1}$, Sudheer Sreedhar $^{1}$, Journey Byland $^{1}$, Zihao Shen $^{1}$, Sergey Y. Savrasov $^{1}$, Valentin Taufour$^{1}$, Vsevolod Ivanov$^{3,4,5}$,  and Inna Vishik $^{1,2}$*}

\affil[1]{\small Department of Physics and Astronomy, University of California, Davis, CA 95616, USA}
\affil[2]{Materials Sciences Division, Lawrence Berkeley National Lab, Berkeley, California 94720, USA}
\affil[3]{Department of Physics, Virginia Tech, Blacksburg, Virginia 24061, USA; vivanov@vt.edu}
\affil[4]{Virginia Tech National Security Institute, Blacksburg, Virginia 24060, USA}
\affil[5]{Virginia Tech Center for Quantum Information Science and Engineering, Blacksburg, Virginia 24061, USA}


\date{}
\maketitle
\vspace{-6ex}

\setcounter{secnumdepth}{0}

The figures in the main text were collected in the following conditions/facilities.  Figure 2 data were collected at SSRL using 36-150eV photons with LH polarized light and approx. 15 x 8 $\mu m$ spot size. Figure 3 data were collected at SSRL.  In Fig 4, in-plane FS maps are taken in the G-X orientation with LH polarized light and kz FS maps are taken both with LH and circularly polarized light.   In Fig. 4, panels a, b, c1-g1, and d2-f2 taken at SSRl beamline 5-2 with 26-118 eV photons. Panels c2 and g2 were taken at ALS beamline 4.0.3 (MERLIN) with 70-104eV photon energies.  Figure 5 data were taken at SSRl beamline 5-2 with 26-118eV photons with LH polarization, and Fig. 6 data collected at SSRL with circularly polarized light. 

\begin{figure}[htb]
    \centering
    \includegraphics[width=0.9\textwidth]{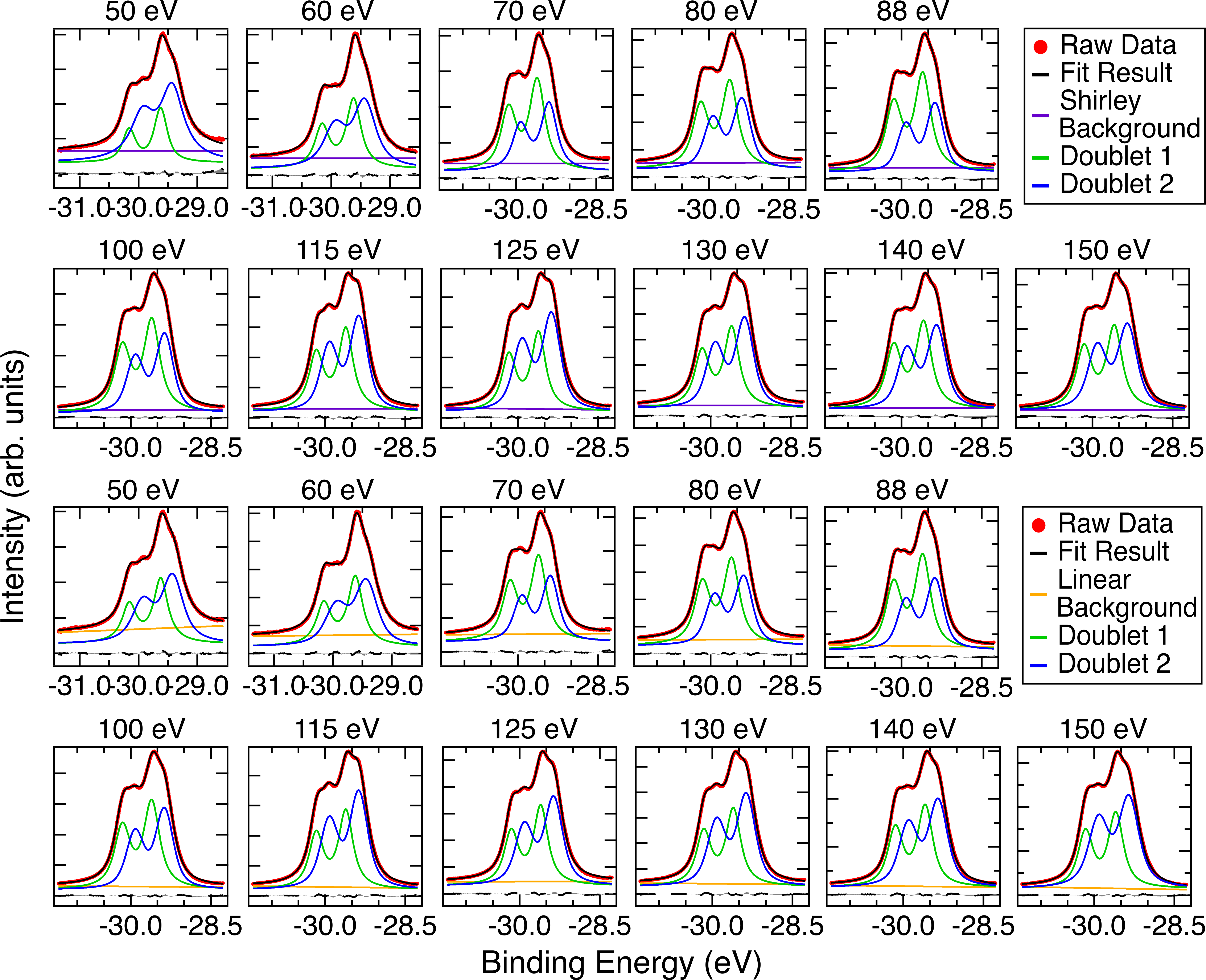}
    \caption{Fit results of Ge-3d core-level spectra with Linear and Shirley background. Top two rows are Shirley Background and bottom two rows are linear background}
    \label{shirley_fits}
\end{figure}

Fig. \ref{shirley_fits} shows fitting of all Ge 3d core levels with 2 different background models. All fittings are done under constraints: 250meV max gaussian width, 2 doublets of the same spectrum share the same spin-orbit energy split and same branch ratio ranging in 1.4 - 1.8.

\begin{figure}[htb]
    \centering
    \includegraphics[width=0.9\textwidth]{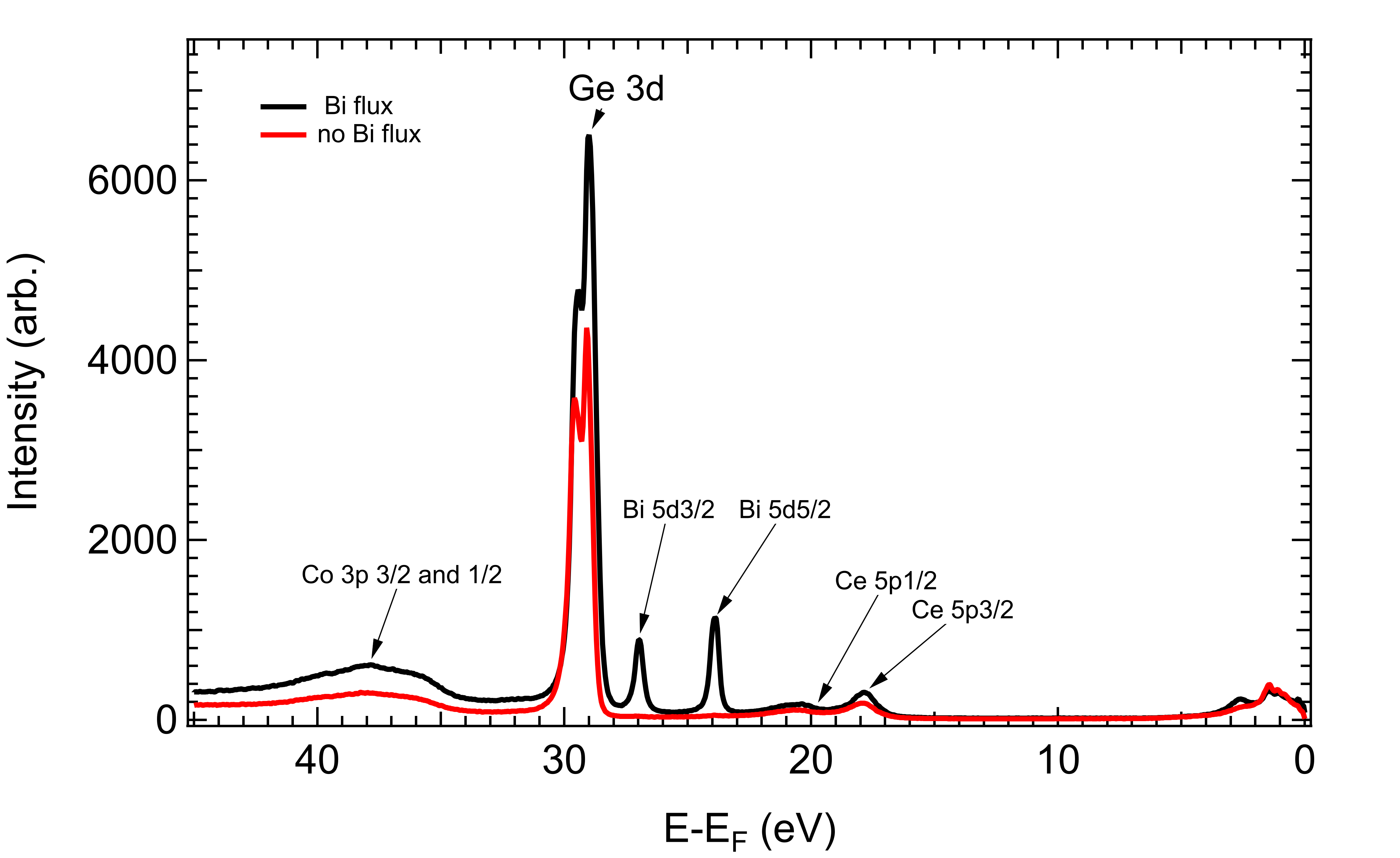}
    \caption{Comparison of survey spectrum with and without Bi flux contamination.}
    \label{BiFlux}
\end{figure}

Fig. \ref{BiFlux} shows example survey XPS spectra with and without Bi-peaks.  The presence of Bi peaks indicates contamination from flux, and these cleaves tend to yield poor ARPES spectra.

\begin{figure}[htb]
    \centering
    \includegraphics[width=0.9\textwidth]{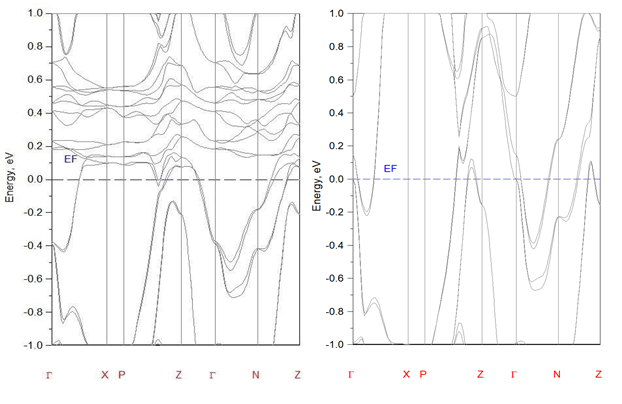}
    \caption{Comparison of LDA calculations with (left) and without (right) inclusion of hybridized Ce electrons.}
    \label{LDA}
\end{figure}

Fig. \ref{LDA} compares LDA calculations including and excluding hybridizing Ce electrons.  The latter (right) shows a shift of dispersing bands to lower binding energy, consistent with experiments.

\begin{figure}[htb]
    \centering
    \includegraphics[width=0.9\textwidth]{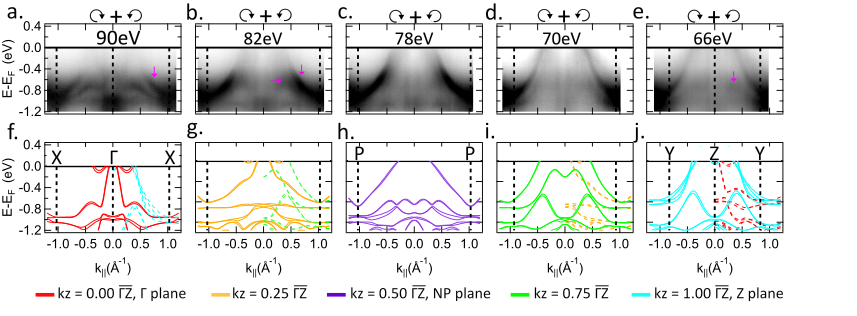}
    \caption{Cuts parallel to $\Gamma-X$ at different $k_z$. (a)-(e) Energy vs momentum cuts at photon energies indicated in each panel. Spectra collected with  RCP+LCP light. Pink arrows point to dispersing feature not directly captured in calculation. (f)-(j) LDA+G calculations from ref \cite{ivanov2021renormalized} corresponding to each data panel above.  Calculations have been shifted to lower binding energy by $180 meV$ for better agreement with data. Vertical dashed lines denote high symmetry points in (f),(h) and (j), and the edge of the BZ at that value of $k_z$ otherwise. Right half of panels (f),(g),(i),(j) shows overlay of calculation from half a BZ away along $k_z$, with color of these dashed lines corresponding to legend labels below}
    \label{GXcuts}
\end{figure}

\begin{figure}[htb]
    \centering
    \includegraphics[width=0.9\textwidth]{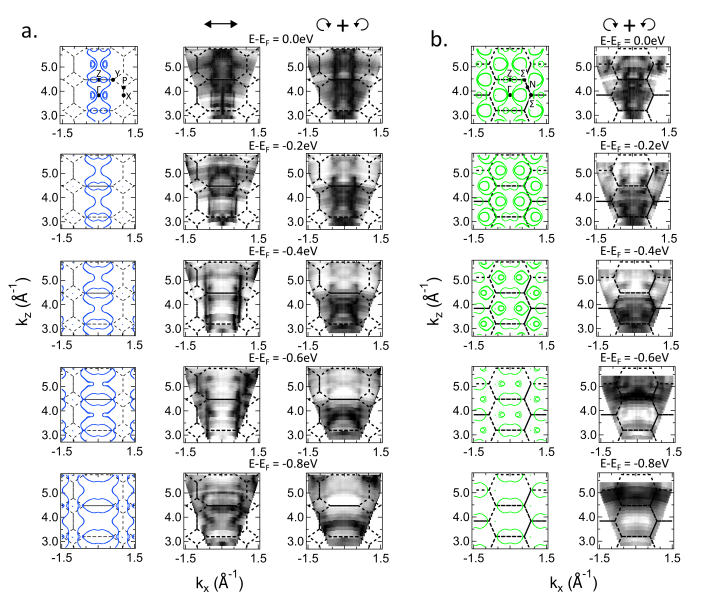}
    \caption{k$_z$, constant energy maps at various binding energies parallel to $\Gamma - X$ (a) and $\Gamma-\Sigma$ (b). Theoretical overlays are plotted in column 1 of (a) and (b) alongside data taken with different polarizations.}
    \label{kz_FSmaps}
\end{figure}

Fig. \ref{GXcuts} shows energy vs momentum cuts parallel to $\Gamma-X$ taken at different photon energies, to complement spectra in $\Gamma-\Sigma$ geometry in the main text.  Pink arrows point to features observed experimentally, but not directly seen in LDA+G for the same cut.  Calculations from half a BZ away in $k_z$ for each cut are also shown.  Fig. \ref{kz_FSmaps} complements Fig. 4 in the main text by showing constant energy maps as a function of $k_z$ at different binding energies and in different geometry.

\begin{figure}[htb]
    \centering
    \includegraphics[width=0.6\textwidth]{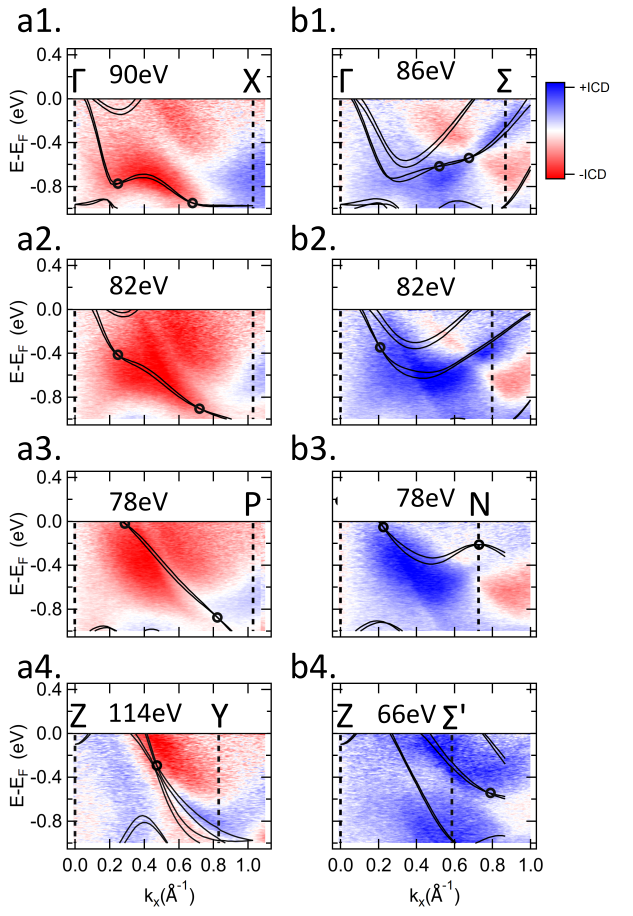}
    \caption{Circular Dichroism (CD), $(I_{RC}-I_{LC})/(I_{RC}+I_{LC})$) of spectra taken at photon energies near 0, 0.25, 0.5, and 1 $\overline{\Gamma Z}$. (a1)-(a4) are spectra taken parallel to $\Gamma-X$ and (b1)-(b4) are taken parallel to $\Gamma-\Sigma$. Black curves are calculated bands dispersions with black circles indicating the positions of topological band crossings which make up the Weyl nodal lines.}
    \label{CDandNLs}
\end{figure}

\begin{figure}[htb]
    \centering
    \includegraphics[width=0.6\textwidth]{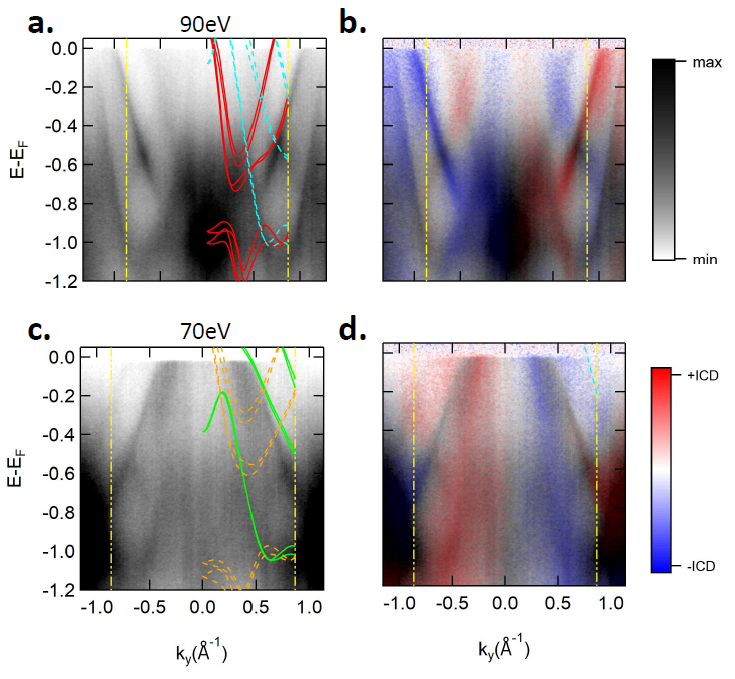}
    \caption{Circular Dichroism in the context of reconstruction band folding, two examples. Left columns: energy vs momentum cut taken with RCP+LCP light. Calculation is for the cut measured (solid) and for half a BZ away in $k_z$ (dashed).  Right columns: same energy vs momentum cut, with circular dichroism image overlaid.}
    \label{CDReconstruction}
\end{figure}

Figs. \ref{CDandNLs} and \ref{CDReconstruction} show circular dichroism energy vs momentum cuts with overlaid band structure to highlight nodal lines (Fig. \ref{CDandNLs}) and unit cell reconstruction (Fig. \ref{CDReconstruction}).   In general, interpretations of CD signals can be quite complex with much work in recent literature describing the systematicity needed to accurately attribute CD signals to chiral features of physical interest \cite{Sidilkover:reexamining2025, Ryu:CD_Au_2017}. In the literature, CD signals have been used as a tool to distinguish surface components in measured band structure \cite{Zabolotnyy:YBCO_surface_dichroism_2007}, and it has also been shown in folded bands whose origin lies elsewhere in the BZ in some charge density wave systems \cite{Seong:ErTe3_2021}. This gives some support for the proposed reconstruction origin of the extra bands in Fig. \ref{CDReconstruction} and Fig. 3 of the main text, as they appear with an opposite CD signal.  There is much work in the recent literature using CD ARPES to measure orbital angular momentum and/or Berry curvature \cite{Cho_2021,Sch_ler_2020,PhysRevLett.122.116402,Unzelmann_2021}; importantly, this is a subtle measurement in which geometric and final state effects need to be ruled out.  Examples of the CD signal in the vicinity of nodal lines in CeGoGe$_3$ is shown in Fig. \ref{CDandNLs}.  We note that in the $\Gamma-\Sigma$ geometry (b-panels), signatures of reconstruction are quite strong and may dominate the CD signal.

\bibliographystyle{unsrtnat}
\bibliography{references}
\newpage

